\documentclass[preprint,showpacs,prb]{revtex4}
\usepackage{graphicx}
\usepackage{color}
\begin{document}
\title{Ionization and Coulomb explosion of Xenon clusters by intense, few-cycle laser pulses}
\author{D. Mathur}
\email{atmol1@tifr.res.in}
\affiliation{Tata Institute of Fundamental Research, 1 Homi Bhabha Road, Mumbai 400 005, India}
\author{F. A. Rajgara} 
\affiliation{Tata Institute of Fundamental Research, 1 Homi Bhabha Road, Mumbai 400 005, India}

\begin{abstract}
Intense, ultrashort pulses  of 800 nm laser light (12 fs, $\sim$4 optical cycles) of peak intensity 5$\times$10$^{14}$ W cm$^{-2}$ have been used to irradiate gas-phase Xe$_n$ clusters ($n$=500-25,000) so as to induce multiple ionization and subsequent Coulomb explosion. Energy distributions of exploding ions are measured in the few-cycle domain that does not allow sufficient time for the cluster to undergo Coulomb-driven expansion. This results in overall dynamics that appear to be significantly different to those in the many-cycle regime. One manifestation is that the maximum ion energies are measured to be much lower than those obtained when longer pulses of the same intensity are used. Ion yields are cluster-size independent but polarization dependent in that they are significantly larger when the polarization is perpendicular to the detection axis than along it. This unexpected behavior is qualitatively rationalized in terms of a spatially anisotropic shielding effect induced by the electronic charge cloud within the cluster.  
\end{abstract}
\maketitle

The physics governing the interaction of intense laser fields with large gas-phase clusters has attracted considerable attention in the course of the last decade or so, and the subject area has developed into a vibrant subset of contemporary research on how matter behaves in strong optical fields. The gas-phase clusters that we refer to comprise a few hundred to several tens of thousands of atoms, most usually rare gas atoms, with sizes such that typical infrared (800 nm wavelength) optical fields  are uniform across the entire cluster. The highly efficient deposition of energy from such optical fields to the cluster is, hence, localized to the cluster. The very earliest work \cite{Boyer,Ditmire} established that the physics of laser-cluster interactions in the strong field regime possesses a number of distinctive features, including (i) qualitative differences in the mechanisms and level of ionization achieved in clusters in comparison to that in atoms and molecules, (ii) the high efficiency of energy absorption by a cluster and its redistribution within the cluster and (iii) upon Coulomb explosion of highly-charged clusters, the acceleration mechanisms that account for production of ions such that they possess energies that exceed the Coulomb limit. Probing these differences has been the key driver for progress in this area of research, and a number of reviews have been published \cite{Saalmann,Krainov,Rostock,Mathur} that provide a cogent overview of different insights that have been gained in recent years. 

A major physics question that remains open after sustained experimental and theoretical work concerns the mechanism by which the optical field's energy is transferred to the cluster. Electrons within the cluster are the direct absorbers of optical energy, most likely through collisional absorption (inverse bremsstrahlung, IB). IB involves the transferring of energy from electrons to other particles in a series of inelastic collisions within the nanoplasma that is created upon initial field-induced inner-ionization of the irradiated cluster. In cluster parlance, `inner ionization' refers to ejection of electrons from the individual Xe-atoms that constitute a cluster in our experiments. Such electrons are quasi-free: they are not bound to any specific Xe-atom but they are spatially constrained within the cluster by the Coulombic charge of the positive ions. It is these spatially constrained electrons that are subjected to heating by the oscillating optical field and the ensuing collisions within the nanoplasma can be thought of as being responsible for the ``bulk temperature" of the plasma. As further inner ionization occurs, the build-up of ionic charge results in the hot nanoplasma undergoing spatial expansion and breaking up in a few picoseconds, yielding energetic electrons, x-rays, as well as ions with kinetic energies that can be as high as a few MeV \cite{Ditmire}, up to five orders of magnitude larger than the energies obtained upon Coulomb explosion of monomeric species like multiply-charged molecules \cite{MathurPhysRep}. 

Mechanisms other than IB that might play a role in energy transfer from the optical field to the cluster are of non-collisional nature, including resonance absorption \cite{Ginzburg} and vacuum (or Brunel) heating \cite{Brunel}. We obviate the need to consider these other processes in the present study by taking recourse to few-cycle laser pulses with peak intensities that lie below 10$^{15}$ W cm$^{-2}$. Our intensity regime allows us to discount resonance absorption. Additionally, our temporal regime allows us to satisfactorily deal with a less-addressed facet of the collisional mechanism, namely the role played by enhanced ionization (EI) as a possible contributor to energy absorption by the cluster. The role of EI in clusters has not been explicitly addressed in any experimental study thus far \cite{Saalmann}. We effectively `switch off' EI: our pulses are too short for significant nuclear motion to occur. 

Hitherto-existing experimental work in this area has overwhelmingly been carried out using infrared laser pulses of 30 fs duration or longer ($>$10 optical cycles). We report here results of experiments in which we use intense pulses of 800 nm wavelength laser light in the {\em few-cycle} domain (12 fs pulse duration, long enough for only $\sim$4 optical cycles). Specifically, we report on energy distributions of ions that are ejected upon Coulomb explosion of highly charged Xe$_n$ clusters ($n$=500-25,000) upon irradiation at 5$\times$10$^{14}$ W cm$^{-2}$ intensity. As already noted, our temporal regime does now allow the irradiated cluster sufficient time to undergo Coulomb-driven expansion and, consequently, the ionic degrees of freedom are effectively `frozen'. This results in overall dynamics that appear to be significantly different to those in the many-cycle regime, one manifestation being that we measure the maximum ion energies in our few-cycle experiments to be much lower than those obtained when longer pulses of the same intensity are used. Ion yields in our experiments are found to be polarization dependent in that they are significantly larger when the polarization is perpendicular to the detection axis than along it. This unexpected behavior is qualitatively rationalized in terms of a spatially anisotropic shielding effect induced by the electronic charge cloud within the cluster. The results that we present in the following throw new light on laser-cluster interactions in the ultrashort, strong-field domain and open opportunities for further work that is qualitatively new. 

Generation of intense few-cycle optical pulses has become important in the contemporary pursuit of attosecond pulses and in studies of light-matter interactions in general. In most laboratories it is the hollow-fiber pulse (HFP) compression technique \cite{Nisoli} that is used to generate few-cycle pulses although an alternate method based on filamentation in rare gases has recently begun to also find utility \cite{filamentation}. It is the latter technique that has been adopted in our laboratory \cite{our_few_cycles} to probe the ionization and fragmentation dynamics of molecules \cite{our_molecules} with four-cycle pulses, and to demonstrate a unimolecular bond rearrangement in H$_2$O on the timescale of a single vibrational period \cite{our_rearrangement}. Generation of gas-phase clusters is by standard techniques: the mean size of our Xe-clusters was varied from Xe$_{500}$ to Xe$_{25,000}$ by controlling the stagnation pressure behind a high-pressure gas nozzle \cite{vinod_asymmetry,vinod2}. Standard time-of-flight methods were used to mass spectrometrically separate the ionic charge states obtained upon cluster explosion. A simple transformation of time-of-flight spectra enabled temporal information to be mapped into energy spectra \cite{vinod2}, typical examples of which are shown in the following.

Figure 1 shows typical energy spectra measured upon irradiation of Xe$_{500}$-Xe$_{25,000}$ clusters by single pulses of 12 fs duration, with peak laser intensity of 5$\times$10$^{14}$ W cm$^{-2}$ in each instance. We note the following facets of these energy distributions that distinguish them from those that are obtained when measurements are made with many-cycle laser pulses of the same intensity \cite{Saalmann}.

\begin{figure}
\includegraphics[width=10cm]{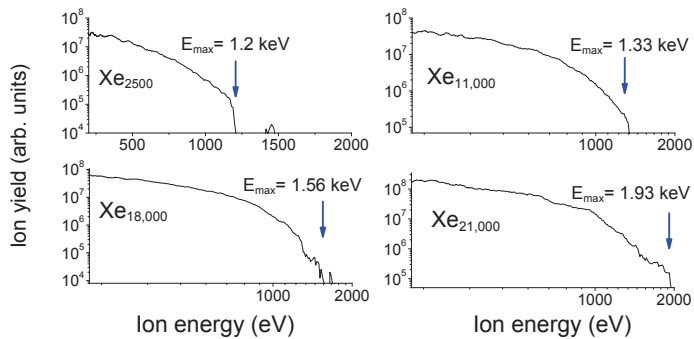}
\caption{Energy distributions of ions obtained upon irradiation of Xe-clusters with 4-cycle (12 fs) pulses of 800 nm laser light of intensity 5$\times$10$^{14}$ W cm$^{-2}$. The laser polarization vector was perpendicular to the ion detection axis. Very much smaller ion yields were obtained when the laser polarization was parallel to the ion detection axis.}
\end{figure}

The shapes of the energy distribution functions are noteworthy in that they differ significantly from shapes that are obtained in longer-pulse experiments. Secondly, the maximum ion energies lie in the relatively narrow range 1.2-1.9 keV as the cluster size varies over a factor of 40: it is clear that there is little dependence of maximum ion energy on cluster size.  The other facet that is unexpected is the strong anisotropy of the ion energy spectrum with respect to the laser polarization vector that we observe: the Xe-ion signal that we detected perpendicular to the laser polarization vector was very much higher than in the parallel direction. This is also in marked contrast to the findings in many-cycle experiments. 

On the first facet noted above, earlier work \cite{vinod_asymmetry,vinod2} has established that ion energy distributions obtained with 100 fs pulses consists of two energy domains. One domain gives rise to maximum values of ion energies that are consistent with what would be expected upon Coulomb explosion of a charged spherical cluster of given size. This component is isotropic with respect to the incident polarization vector. Additionally, there is a second, higher energy ion component whose maximum energies extend beyond the Coulomb limit. This high energy component exhibits anisotropy with respect to the laser polarization vector \cite{vinod_asymmetry,vinod2}. The two components are separated in measured energy distribution functions by a knee-like inflection that is missing in the few-cycle data shown in Fig. 1. The angular anisotropy experimentally observed with 100 fs pulses was subsequently theoretically rationalized \cite{Saalmann} to be a consequence of polarization-induced effects that arise from the phase difference between the optical field and the oscillations of the negatively-charged cloud generated within the cluster by inner-ionized electrons, resulting in a periodic ``flipping" of charges at the two poles of an initially spherical cluster \cite{vinod_asymmetry,vinod2}, with the poles being aligned along the laser's $E$-vector. However, such charge flipping cannot be expected to be of consequence when cluster irradiation is by a 4-cycle pulse as there would be insufficient time for significant phase difference to accumulate. The asymmetric ion and electron emission that has been reported upon cluster disassembly \cite{vinod_asymmetry,vinod2} in the many-cycle temporal regime would, consequently, not be expected when few-cycle pulses are employed. 

We now take note of the second facet of our results, namely the size independence of maximum ion energy that is observed in our 4-cycle experiments. This result makes clear that there is insufficient time for a large enough fraction of the available electrons within Xe$_{500}$-Xe$_{25,000}$ clusters to be inner-ionized. If a large enough fraction were to be inner-ionized, as is the case with, say, 100 fs pulses, then the size of the cluster would matter: larger clusters would have a larger number of inner-ionized electrons which would absorb more energy from the optical field and drive Coulomb expansion of the cluster more effectively; in turn, this would give rise to more energetic ions being emitted upon Coulomb explosion. If, on the other hand, the pulse duration is so short that only a very small fraction of available electrons are inner-ionized, the size-dependence is considerably scaled down, as seems to be consistent with observation (Fig. 1). It would clearly be of interest to carry out theoretical simulations of the scenario that we have postulated and it is hoped that our experiments stimulate this.

The third facet of our results, namely the very strong anisotropy of the ion energy spectrum with respect to the laser polarization vector, is also likely to be related to the charge cloud of inner-ionized electrons. This charge cloud will be spatially asymmetric in that electrons will follow the laser polarization vector. Consequently, the electronic charge will be expected to be more dense along the polarization direction than perpendicular to it. The electron charge cloud serves to shield ions within the core of the cluster in such fashion that the extent of shielding is spatially inhomogeneous: the polar regions of the spherical cluster (in the direction of the laser polarization vector) are much more effectively shielded than the equatorial regions. Consequently, when the ionic charge cloud in the core builds up sufficiently to result in a Coulomb explosion, the products of this explosion that emerge out of the cluster manifest themselves more prominently in the equatorial region than in the polar regions. This is precisely what is observed in our experiments, with a very significant diminishing of ion yields in the laser polarization direction parallel to the detection axis compared to what is obtained in the orthogonal direction. In long-pulse experiments, such shielding is of little consequence as the charge-flipping process referred to above dominates the ion dynamics and leads to the opposite type of asymmetric emission of Coulomb explosion products, more along the polarization direction than perpendicular to it. 

\begin{figure}
\includegraphics[width=6cm]{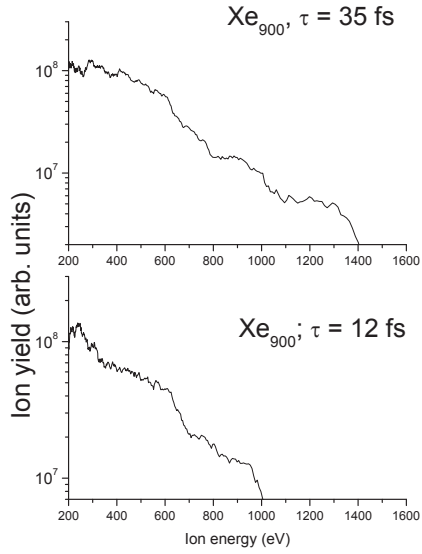}
\caption{Ion energy spectra measured using pulses of 35 fs (top panel) and 12 fs (lower panel) duration, both of peak intensity 5$\times$10$^{14}$ W cm$^{-2}$.}
\end{figure}

Apart from energy spectra, spatially asymmetric shielding will also be expected to manifest itself in ion yields. This is, indeed, the case, as has been reported recently in the case of Ar-clusters \cite{arxiv}. 

One consequence of the postulates we have made above is amenable to immediate experimental test: laser pulse duration ought to have a significant effect on the extent of electronic charge cloud shielding and ought, therefore, to have an effect on the ion energy distributions. We have carried out experiments on Xe-clusters using both 12 fs and 35 fs pulses and results for Xe$_{900}$ clusters are shown in Fig. 2.  The pulse energy was appropriately compensated in these two measurements to ensure that the peak intensity experienced by Xe$_{900}$ clusters was the same. The maximum energy obtained with 35 fs pulses is seen to be significantly larger than in the case of 12 fs pulses. It is of interest to note that a nonlinear theoretical treatment made by Breizman {\it et al.} \cite{Breizman} on laser-cluster interactions with ultrashort pulses predicts that the ion energy should scale as $\tau^2$, where $\tau$ denotes the laser pulse duration. It is clear from data that we have accumulated in our experiments that this scaling is not observed. More elaborate theoretical work is, clearly, necessary.

\acknowledgements
We gratefully acknowledge the Department of Science and Technology, Government of India for financial support.

\end{document}